\documentclass[conference]{IEEEtran}
\IEEEoverridecommandlockouts
\usepackage[letterpaper, top=0.78in, bottom=1.08in, left=0.625in, right=0.625in]{geometry} 

\usepackage[mathscr]{eucal}
\usepackage[cmex10]{amsmath}
\usepackage{epsfig,epsf,psfrag}
\usepackage{amssymb,amsthm,amsfonts,latexsym}
\usepackage{xcolor,url,overpic}
\usepackage{caption}
\usepackage{subcaption}
\usepackage{array}
\usepackage{verbatim}
\usepackage{bm}
\usepackage{bbm}
\usepackage{dsfont}
\usepackage{algorithmic}
\usepackage{algorithm}
\usepackage{verbatim}
\usepackage{textcomp}
\usepackage{mathrsfs}
\usepackage{epstopdf}
\usepackage{booktabs} 
\usepackage{thm-restate}
\usepackage{tikz}

\newcommand{\openone}{\leavevmode\hbox{\small1\normalsize\kern-.33em1}}

\catcode`~=11 \def\UrlSpecials{\do\~{\kern -.15em\lower .7ex\hbox{~}\kern .04em}} \catcode`~=13 

\allowdisplaybreaks[1]

\newcommand{\nn}{\nonumber}

\newcommand{\calA}{\mathcal{A}}
\newcommand{\calB}{\mathcal{B}}

\newcommand{\calG}{\mathcal{G}}

\newcommand{\calP}{\mathcal{P}}

\newcommand{\calR}{\mathcal{R}}

\newcommand{\calV}{\mathcal{V}}

\newcommand{\calX}{\mathcal{X}}

\newcommand{\calZ}{\mathcal{Z}}

\newcommand{\rmc}{\mathrm{c}}

\newcommand{\rmH}{\mathrm{H}}

\newcommand{\bbE}{\mathbb{E}}
\newcommand{\bbF}{\mathbb{F}}

\newcommand{\bbP}{\mathbb{P}}

\newcommand{\bbR}{\mathbb{R}}

\newcommand{\bbZ}{\mathbb{Z}}

\DeclareMathAlphabet{\mathbsf}{OT1}{cmss}{bx}{n}
\DeclareMathAlphabet{\mathssf}{OT1}{cmss}{m}{sl}

\DeclareSymbolFont{bsfletters}{OT1}{cmss}{bx}{n}  
\DeclareSymbolFont{ssfletters}{OT1}{cmss}{m}{n}
\DeclareMathSymbol{\bsfGamma}{0}{bsfletters}{'000}
\DeclareMathSymbol{\ssfGamma}{0}{ssfletters}{'000}
\DeclareMathSymbol{\bsfDelta}{0}{bsfletters}{'001}
\DeclareMathSymbol{\ssfDelta}{0}{ssfletters}{'001}
\DeclareMathSymbol{\bsfTheta}{0}{bsfletters}{'002}
\DeclareMathSymbol{\ssfTheta}{0}{ssfletters}{'002}
\DeclareMathSymbol{\bsfLambda}{0}{bsfletters}{'003}
\DeclareMathSymbol{\ssfLambda}{0}{ssfletters}{'003}
\DeclareMathSymbol{\bsfXi}{0}{bsfletters}{'004}
\DeclareMathSymbol{\ssfXi}{0}{ssfletters}{'004}
\DeclareMathSymbol{\bsfPi}{0}{bsfletters}{'005}
\DeclareMathSymbol{\ssfPi}{0}{ssfletters}{'005}
\DeclareMathSymbol{\bsfSigma}{0}{bsfletters}{'006}
\DeclareMathSymbol{\ssfSigma}{0}{ssfletters}{'006}
\DeclareMathSymbol{\bsfUpsilon}{0}{bsfletters}{'007}
\DeclareMathSymbol{\ssfUpsilon}{0}{ssfletters}{'007}
\DeclareMathSymbol{\bsfPhi}{0}{bsfletters}{'010}
\DeclareMathSymbol{\ssfPhi}{0}{ssfletters}{'010}
\DeclareMathSymbol{\bsfPsi}{0}{bsfletters}{'011}
\DeclareMathSymbol{\ssfPsi}{0}{ssfletters}{'011}
\DeclareMathSymbol{\bsfOmega}{0}{bsfletters}{'012}
\DeclareMathSymbol{\ssfOmega}{0}{ssfletters}{'012}

\newcommand{\hatM}{\hat{M}}

\newcommand{\iid}{i.i.d.\ }

\newcommand{\dotleq}{\stackrel{.}{\leq}}

\newcommand{\eqb}{\stackrel{(b)}{=}}

\DeclareMathOperator*{\argmax}{arg\,max}
\DeclareMathOperator*{\argmin}{arg\,min}

\usepackage{thmtools, thm-restate}
\newtheorem{theorem}{Theorem} 
\newtheorem{lemma}[theorem]{Lemma}

\newtheorem{definition}{Definition} 
\newtheorem{example}{Example} 
 
\newtheorem{remark}{Remark}

\newtheorem{assumption}{Assumption}

\usepackage{cite}
\usepackage{amsmath,amssymb,amsfonts}
\usepackage{algorithmic}
\usepackage{graphicx}
\usepackage{textcomp}
\usepackage{xcolor}
\def\BibTeX{{\rm B\kern-.05em{\sc i\kern-.025em b}\kern-.08em
    T\kern-.1667em\lower.7ex\hbox{E}\kern-.125emX}}

\usepackage{mathtools}
\usepackage{cases}
\usepackage{mdframed}
\usepackage[shortlabels]{enumitem}
\usepackage{extarrows}
\usepackage{tikz}
\usetikzlibrary{positioning}
\usetikzlibrary{calc}

\mathtoolsset{showonlyrefs}

\newcommand{\KL}{\mathsf{KL}}
\newcommand{\TV}{\mathsf{TV}}

\mathchardef\mhyphen="2D

\newcommand{\sD}{\mathsf{D}}

\newcommand{\sH}{\mathsf{H}}

\newcommand{\sI}{\mathsf{I}}

\begin{document}

\title{Distributional Information Embedding:\\ A Framework for Multi-bit Watermarking} 

\author{%
    \IEEEauthorblockN{
    \textbf{Haiyun~He}\IEEEauthorrefmark{1}, 
    \textbf{Yepeng~Liu}\IEEEauthorrefmark{2}, 
    \textbf{Ziqiao~Wang}\IEEEauthorrefmark{3},
    \textbf{Yongyi~Mao}\IEEEauthorrefmark{4}, 
    and \textbf{Yuheng~Bu}\IEEEauthorrefmark{2}}
   \IEEEauthorblockA{\IEEEauthorrefmark{1}%
                     Center for Applied Mathematics, 
                     Cornell University, 
                     Ithaca, NY, USA}
   \IEEEauthorblockA{\IEEEauthorrefmark{2}%
                     Department of CS, 
                     University of California, Santa Barbara, 
                     Santa Barbara, CA, USA}
    \IEEEauthorblockA{\IEEEauthorrefmark{3}%
                     School of Computer Science and Technology, 
                     Tongji University, 
                     Shanghai, China}
    \IEEEauthorblockA{\IEEEauthorrefmark{4}%
                     Schoolf of EECS, 
                     University of Ottawa, Ottawa, ON, Canada}
                      
                     \begin{small}{Emails: hh743@cornell.edu, \{yepengliu, buyuheng\}@ucsb.edu, ziqiaowang@tongji.edu.cn, ymao@uottawa.ca}\end{small} 
                     \vspace{-4ex}
}

\maketitle

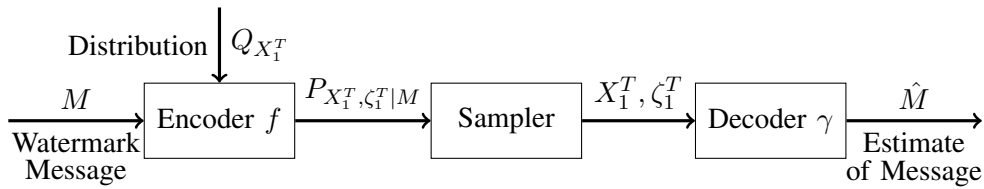
\begin{figure*}[t]
\centering
\begin{tikzpicture}



    \node (start) at (2.5, 1) {};

    \tikzset{
        mybox/.style={
            rectangle, 
            draw, 
            fill=white, 
            minimum height=1cm, 
            minimum width=2cm, 
            align=center
        }
    }

        \node[mybox] (box1) at (start) {Encoder $f$};
        \node[mybox, right=1.8 of box1] (box2) {Sampler};
        \node[mybox, right=1.5 of box2] (box3) {Decoder $\gamma$};

        \draw[<-, very thick] (box1.north) -- ++(0,0.7) node[midway, left] {Distribution} node[midway, right] {$Q_{X^T}$};

        \draw[<-, very thick] (box1.west) -- ++(-1.8, 0) node[midway, above] {$M$} node[midway, below] {\shortstack{Watermark \\ Message}};

        \draw[->, very thick] (box1.east) -- ++(1.8, 0) node[midway, above] {$P_{X^T,\zeta^T|M}$};

        \draw[->, very thick] (box2.east) -- ++(1.5, 0) node[midway, above] {$X^T,\zeta^T$};
        
        \draw[->, very thick] (box3.east) -- ++(1.8, 0) node[midway, above] {$\hat{M}$} node[midway, below] {\shortstack{Estimate \\ of Message}};
\end{tikzpicture}
\caption{Illustration of multi-bit watermarking as distributional information embedding with side information.}
\label{Fig:WM system}
\vspace{-1em}
\end{figure*}

\begin{abstract}
This paper introduces a novel problem, distributional information embedding, motivated by the practical demands of multi-bit watermarking for large language models (LLMs). Unlike traditional information embedding, which embeds information into a pre-existing host signal, LLM watermarking actively controls the text generation process—adjusting the token distribution—to embed a detectable signal. We develop an information-theoretic framework to analyze this distributional information embedding problem, characterizing the fundamental trade-offs among three critical performance metrics: text quality, detectability, and information rate. In the asymptotic regime, we demonstrate that the maximum achievable rate with vanishing error corresponds to the entropy of the LLM's output distribution and increases with higher allowable distortion. We also characterize the optimal watermarking scheme to achieve this rate. Extending the analysis to the finite-token case with non-\iid tokens, we identify schemes that maximize detection probability while adhering to constraints on false alarm and distortion. 



\end{abstract}


\section{Introduction}

The rapid advancement of Large Language Models (LLMs)~\cite{touvron2023llama,jiang2023mistral} is revolutionizing numerous fields but also raises concerns about misuse, such as spreading disinformation, creating fake news, and enabling academic dishonesty. The growing prevalence and quality of AI-generated text make it challenging to \emph{distinguish it from human-written content}.

A promising solution is to \emph{actively} embed detectable signals into LLM-generated text, i.e., watermarks, which enable provable detection of AI-generated content.
Despite recent advances in watermarking algorithms for LLM \cite{gumbel2023,kirchenbauer2023watermark, kuditipudi2023robust, zhao2023provable, liu2024adaptive}, they suffer from significant limitations, for example, many algorithms are heuristically designed where watermark detectability is ensured by introducing noticeable alterations to the generated content that degrade the output quality. 

Additionally, most watermarking schemes are ``zero-bit'' schemes, designed solely to distinguish AI-generated text from human-written content without embedding any additional information. As incorporating meta-information—such as the model's name, version, and generation time—is increasingly important for forensic analysis of LLM misuse, some multi-bit watermarking algorithms~\cite{yoo2023robust,yoo2024advancing,qu2024provably} have been developed recently. However, these approaches remain heuristic and have a low information embedding rate, with current methods unable to support messages longer than a few bits \cite{zhao2024sok}.

Therefore, a \emph{principled theoretical framework} is needed to analyze the fundamental trade-offs among key performance metrics in multi-bit LLM watermarking. These metrics include: (1) \textbf{Text quality}: ensuring that the watermarked text generated by LLMs maintains a quality comparable to unwatermarked text; (2) \textbf{Detectability}: the probability of missed detection and decoding errors; and (3) \textbf{Information rate}: the rate at which information can be embedded and reliably  recovered. 

Information theory has a long-standing history of guiding the design of digital watermarking, dating back to the early 00s \cite{moulin2000information,merhav2000random,moulin2001role,steinberg2001identification,cohen2002gaussian}, within the broader framework of the information embedding problem \cite{chen2001quantization,barron2003duality,moulin2003information,eggers2003scalar, 4529296, harmsen2009capacity}. As we will demonstrate, watermarking in LLMs introduces a novel form of such a problem, which we term \emph{distributional information embedding}. Unlike traditional information embedding, which focuses on reliably embedding information into a pre-existing host signal while minimizing distortion, LLM watermarking actively controls the generation process—the token distribution—to embed a detectable signal while preserving the original distribution. In other words, traditional information embedding is like writing on dirty paper \cite{costa1983writing}, where the challenge is to convey the message clearly despite the interference from pre-existing marks. In contrast, LLM watermarking resembles generating dirty paper in real time, embedding the message into the very process that creates the marks.
This fundamental difference reshapes the problem and introduces novel challenges.

In this paper, we present an information-theoretic analysis of a distributional information embedding problem motivated by multi-bit LLM watermarking. Our goal is to design the watermarking scheme by jointly optimizing the encoder and decoder. The system must distinguish human-written text from AI-generated text while ensuring reliable recovery of the embedded information. All of this must be achieved within a specified distortion constraint to preserve text quality.
Our contribution includes:
\begin{itemize}
[leftmargin=*,topsep=0.2em,itemsep=0.2em]
    \item \textbf{Asymptotic analysis in the independent and identically distributed (i.i.d.) case}: 
    We demonstrate that the maximum information rate with vanishing detection error probability corresponds to the entropy of the LLM's output distribution and increases with higher allowable distortion. Furthermore, we characterize the asymptotically optimal watermarking scheme that achieves this rate.
    \item \textbf{Finite token length analysis in the non-i.i.d.~case}: We extend the asymptotic analysis to the practical scenario with a finite token length, aiming to maximize the detection accuracy while satisfying both a worst-case false alarm probability constraint and a distortion constraint. Furthermore, we characterize the minimum achievable detection error and identify the optimal watermarking scheme and decoder for this setting.
\end{itemize}

\vspace{-0.5em}
\section{Problem Formulation}


\paragraph{Distributional Information Embedding with Side Information}
Consider a length-$T$ data sequence $X^T$ generated from a joint distribution $Q_{X^T}\in\calP(\calX^T)$, where $\calP(\calX^T)$ denotes the probability simplex in $\calX^T$. For simplicity, we ignore the potential auto-regressive structure of $Q_{X^T}$ in the current analysis. 
In the generation process, a message $M$ drawn from $[m]\coloneqq\{1,\ldots,m\}$ needs to be embedded in the data sequence by constructing a dependence structure between $X^T$ and an auxiliary random sequence $\zeta^T$ with alphabet $\calZ^T$, which serves as side information available to the decoder.

For example, the joint distribution $Q_{X^T}$ can be viewed as the output distribution of an LLM for a length-$T$ token sequence $X^T$.  Most LLM watermarking schemes adopt this distributional information embedding with side information framework. An example of a zero-bit watermarking scheme is provided below.

\begin{example}[Existing watermarking schemes as special cases]\label{Ex: watermark schemes}
    In the Green-Red List watermarking scheme \cite{kirchenbauer2023watermark}, at each position $t$, the token vocabulary $\calX$ is randomly split into a green list $\calG$ and a red list $\calR$, with $|\calG|=\rho|\calX|$. This split is represented by a $|\calX|$-dimensional binary auxiliary variable $\zeta_t$, indexed by $x \in \calX$, where $\zeta_t(x)=1$ means $x\in\calG$; otherwise, $x\in\calR$.  
    The watermarking scheme is as follows:
    \begin{enumerate}[noitemsep,topsep=0pt,parsep=0pt,partopsep=0pt, leftmargin=*,label=--]
        \item Compute a hash of the previous token $X_{t-1}$ using a hash function $hash:\calX\times \bbR\to\bbR$ and a shared secret key: $hash(X_{t-1},\mathrm{key})$.
        \item Use $hash(X_{t-1},\mathrm{key})$ as a seed to uniformly sample the auxiliary variable $\zeta_t$ from the set $\{\zeta\in\{0,1\}^{|\calX|}: \|\zeta\|_1=\rho|\calX|\}$ to construct the green list $\calG$.
        \item Sample $X_t$ from the modified token-generating distribution which increases the logit of tokens in $\calG$ by $\delta>0$:
    \begin{equation}
        P_{X_t|x^{t-1},\zeta_t}(x)= \frac{Q_{X_t|x^{t-1}}(x)\exp(\delta\cdot\mathbbm{1}\{\zeta_t(x)=1\})}{\sum_{x\in\calV}Q_{X_t|x^{t-1}}(x)\exp(\delta\cdot\mathbbm{1}\{\zeta_t(x)=1\})}.
    \end{equation}
    \end{enumerate}
    
    
    More examples of several other watermarking schemes, e.g., Gumbel-Max \cite{gumbel2023}, EXP-Edit \cite{kuditipudi2023robust} and text-adaptive watermark \cite{liu2024adaptive}, is provided in \cite[Appendix A]{he2024universally}.
\end{example}





In this paper, we focus on studying one usage scenario within this framework: multi-bit watermarking. Below, we formulate the  multi-bit watermarking problem during data generation as a distributional information embedding problem with side information, as illustrated in Figure \ref{Fig:WM system}. 

\begin{definition}[Multi-bit Watermarking] A watermarking system is an encoder/decoder pair $(f, \gamma)$.
The encoder $f:[m]\times \calP(\calX^T) \to \calP(\calX^T\times\calZ^T|[m])$ inputs a watermark message $M$ drawn from the index set $[m]$ and the data generation distribution $Q_{X^T}$, outputting a joint distribution $P_{X^T,\zeta^T|M}$ that creates dependence between the generated data and auxiliary random sequence $\zeta^T$. The decoder receives $(X^T,\zeta^T)$ sampled from $P_{X^T,\zeta^T|M}$, and guesses the message $M$ with decoder $\gamma:\calX^T\times\calZ^T \to [0\!:\!m]$, i.e., $\hat{M}=\gamma(X^T,\zeta^T)$. If $\hat{M}=0$, the  sequence $X^T$ is decoded as unwatermarked; if $\hat{M}\in [m]$, $X^T$ is decoded as watermarked with   message $\hat{M}$. This system defines an $(m,T)$ watermarking scheme with an information rate $R\coloneqq \log m / T$.
\end{definition}

 
%

Note that the watermarked sequence is generated from $P_{X^T}$ (induced by the encoder $f$) instead of the original $Q_{X^T}$. To measure the \emph{distortion level} of a watermarking scheme, 
we use the divergence between these two distributions. 
\begin{definition}[$d$-Distorted Watermarking]\label{Def: distortion}
    A watermarking encoder $f$ is \emph{$d$-distorted} with respect to the distortion $\sD$, if for any $M \in [m]$ and $Q_{X^T} \in \calP(\calX^T)$, the marginal distribution of the output $P_{X^T,\zeta^T|M}$ satisfies $\sD(P_{X^T|M}, Q_{X^T})\leq d$. 
\end{definition}
Here, $\sD$ can be any divergence. Common examples of such divergences include total variation, KL divergence, and Wasserstein distance. For $d=0$, the watermarking scheme is called \emph{distortion-free}.

Moreover, to ensure the secrecy of the embedded message, we assume that the watermarked  sequence $X^T$ should be indistinguishable for any embedded message $M$,
provided the auxiliary sequence is unknown. A distortion-free watermarking scheme satisfies this condition, as it ensures $P_{X^T|M=j}=Q_{X^T}$, for all $j\in[m]$. Additionally, the auxiliary sequence itself should not reveal any information about the message. Otherwise, the message $M$ can be transmitted directly via the dependence between $\zeta$ and $M$, bypassing the need for the generated text.

\begin{assumption}[Secrecy of Embedded Message] \label{Ass: independence}
    The encoder $f$ must ensure that both $X^T$ and $\zeta^T$ are statistically independent of the embedded message $M$. 
\end{assumption}
Under this assumption, the embedded message $M$ cannot  be decoded with only $\smash{X^T}$ or $\smash{\zeta^T}$ and $\sI(M;X^T,\zeta^T)=\sI(M;X^T|\zeta^T)=\sI(M;\zeta^T|X^T)$.  To detect if $X^T$ is watermarked, the decoder must exploit the auxiliary sequence $\zeta^T$. This corresponds to decoding with side information.

\paragraph{Watermark Detection and Decoding} Under our framework, if the token sequence $X^T$ is unwatermarked, it is independent of  $\zeta^T$; otherwise, $(X^T,\zeta^T)$ is jointly distributed according to one of the $m$ distributions $\{P_{X^T,\zeta^T|M=j}\}_{j=1}^m$. Thus, detecting and decoding the watermark message $M$ boils down to the $(m+1)$-ary hypothesis testing:
\begin{itemize}[leftmargin=*]
    \item $\rmH_0$: $X^T$ is generated by a human, i.e., $(X^T,\zeta^T)\sim \bbP_0\coloneqq P_{X^T,\zeta^T|M=0}= Q_{X^T}\otimes P_{\zeta^T}$;
    \item $\rmH_j, \forall j\in [m]$: $X^T$ is generated by a watermarked LLM and embedded with message $j$, $(X^T,\zeta^T)\sim \bbP_j\coloneqq P_{X^T,\zeta^T|M=j}$.
\end{itemize}

Detection performance is measured by the $j$-th error probability: for any $j\in[0:m]$,
\begin{equation}
    \beta_j(\gamma,P_{X^T,\zeta^T|M=j})\coloneqq\bbP_j(\gamma(X^T,\zeta^T)\ne j ).
\end{equation}
Note that for $j\ne 0$, $\beta_j(\gamma,P_{X^T,\zeta^T|M=j})=\bbP_j(\gamma(X^T,\zeta^T)=0)+\bbP_j(\gamma(X^T,\zeta^T)\in [m]\backslash j)$ is the sum of miss detection error and miss decoding error. For $j=0$, $\beta_0(\gamma,Q_{X^T}\otimes P_{\zeta^T})$ is the false alarm error. Since human-generated texts can vary widely, we aim to control the \emph{worst-case} false alarm error $\sup_{Q_{X^T}}\beta_0(\gamma,Q_{X^T}\otimes P_{\zeta^T})$ at a given $\alpha\in(0,1)$. 

Our design objective is then three-fold:  1) maximizing the information rate $R$, 2) ensuring the distortion remains bounded by $d$, and 3) minimizing $\beta_j(\gamma,P_{X^T,\zeta^T|M=j})$ for all $j\in[m]$ while the \emph{worst-case} false alarm error $\sup_{Q_{X^T}}\beta_0(\gamma,Q_{X^T}\otimes P_{\zeta^T})$ is controlled.

\section{Asymptotic Results with IID Tokens}
In this section, we begin with an asymptotic analysis by letting the length of tokens $T \to \infty$ for the \iid case to build intuition for the optimal design of the watermarking scheme.

Suppose $X_1,\ldots,X_T$ are \iid with an identical distribution $P_X$, and $\zeta_1,\ldots,\zeta_T$ are \iid with $P_\zeta$. Under each $\rmH_j$, $(X_1,\zeta_1),\ldots,(X_T,\zeta_T)$ are conditionally \iid with distribution $P_{X,\zeta|M=j}$. Specifically, $P_{X,\zeta|M=0}=Q_{X}\otimes P_{\zeta}$. Additionally, we assume a uniform prior distribution of message $M$ on $[m]$.

\subsection{Converse Result}

We first analyze the maximum information rate that an $(m,T)$ watermarking scheme can achieve with vanishing decoding error $\Pr(\hatM \ne M)=\frac{1}{m}\sum_{j=1}^m\beta_j(\gamma,P_{X^T,\zeta^T|M=j})$.


\begin{lemma}[Best Achievable Information Rate]\label{Lem: best rate}
    Given any $Q_X$, $P_X$ satisfying $\sD(P_X^T,Q_X^T)\leq d$, and $\{P_{X,\zeta|M=i}\}_{i=0}^m$,
   if the decoding error $\Pr(\hatM \ne M)\to 0 $ as $T\to\infty$, the information rate of this   $d$-distorted $(m,T)$ watermarking scheme is upper bounded by
   \begin{align}
       R\leq  \sH(P_X)\leq \sup_{P_X:\sD(P_X^T,Q_X^T)\leq d}\sH(P_X),
   \end{align}
   where the last bound holds for all $d$-distorted $(m,T)$ watermarking schemes for the LLM $Q_{X^T}$.
\end{lemma}
 The proof is provided in \cite[Appendix A]{NewMultibitsupp}. Lemma \ref{Lem: best rate} shows that it is impossible for a distortion-free watermarking to embed more than approximately $2^{T\sH(Q_X)}$ messages in a length-$T$ \iid token sequence while achieving vanishing $j$-th error for all $j\in[m]$, regardless of the false alarm probability. As the distortion $d$ increases, a $d$-distorted watermarking can trade off text quality to achieve a higher rate. Note that when $d=0$, this maximum information rate coincides with the steganography capacity characterized in classical works  \cite{4418489,harmsen2009capacity} when there is no adversary and noise.

\subsection{Achievability Result}

Next, we aim to identify the asymptotically optimal watermarking scheme that can achieve vanishing $j$-th errors and the maximum watermarking rate, while ensuring the false alarm error below $\alpha$.

To develop intuition for the optimal design, we first present an upper bound for the $j$-th error exponent
under \iid assumptions.
Specifically, we extend \cite[Lemma 11.8.1]{thomas2006elements} to our $(m+1)$-ary hypothesis testing setting. 


\begin{lemma}[Upper Bound for the $j$-th Error Exponent]\label{Lem:LB for j-th error}
Fix any $j\in[m]$, $\epsilon\in(0,1/2)$ and any set of distributions $\{\bbP_i\}_{i\in[0:m]\backslash \{j\}}$.
Let $\{\calB_{T,i}\}_{i\in[0:m]\backslash \{j\}} \subset \calX^T\times \calZ^T$ be any collection of $m$ disjoint sets of sequences $((x_i,\zeta_i))_{i=1}^T$ such that $\bbP_i(\calB_{T,i})\geq 1-\epsilon$. Let $\calB_{T,j}=(\cup_{i\in[0:m]\backslash \{j\}}\calB_{T,i})^\rmc$. For any $\bbP_j$ such that $\max\limits_{i\in [0:m]\backslash \{j\}}\sD_\KL(P_{X,\zeta|M=i}\|P_{X,\zeta|M=j})<\infty$, 
\begin{align}
   &\quad -\frac{ \log \bbP_j(\calB_{T,j}^\rmc)}{T} \leq E_j^*+\epsilon-\frac{\log(m(1-2\epsilon))}{T},\\
   &\text{where } \!E_j^*\hspace{-3pt}\coloneqq \hspace{-8pt}\max_{\substack{ P_X: \sD(P_X^T,Q_X^T)\leq d}}\min_{i\in[0:m]\backslash \{j\}}\hspace{-5pt}\sD_\KL(P_{X\!,\zeta|M\!=\!i}\|P_{X\!,\zeta|M\!=\!j}).
\end{align}
\end{lemma}
\vspace{-1em}
The proof is provided in  \cite[Appendix B]{NewMultibitsupp}. Lemma \ref{Lem:LB for j-th error} shows that given any watermarking scheme $(\bbP_0,\ldots,\bbP_m)$, the minimum achievable $j$-th error probability for all decoders decays exponentially with the rate $E_j^*$, while other errors are controlled below $\epsilon$. Furthermore, the error exponent depends on the distortion level $d$ (cf.\ Definition \ref{Def: distortion}), which increases as $d$ increases. If the distortion metric is set as $\sD_\KL(Q_X^T\|P_X^T)$,  the rate is further upper bounded by $ \sD_\KL( P_{\zeta}\|P_{\zeta|X,M=j}|Q_X)+d$.


Inspired by Lemma \ref{Lem:LB for j-th error}, we can design the joint distributions $(P_{X,\zeta|M=i})_{i=0}^m$ by maximizing the error exponent $E_j^*$. In this way, the $j$-th error probability decays exponentially to $0$ at the fastest rate.  
One solution is to make the masses of $P_{X,\zeta|M=i}$ and $P_{X,\zeta|M=j}$ concentrated at different locations for $i\ne j$, which leads to $\sD_\KL(P_{X,\zeta|M=i}\|P_{X,\zeta|M=j})\to\infty$. This hints that the optimal joint distribution produced by the encoder $f$ should almost deterministically map $(X^T,\zeta^T)$ to a message $M$. Based on this intuition, we construct the asymptotically jointly optimal encoder/decoder pair in the watermarking scheme.

\begin{figure*}[ht]
    \centering
    \includegraphics[trim=0 2 0 5, clip, width=.57\linewidth]{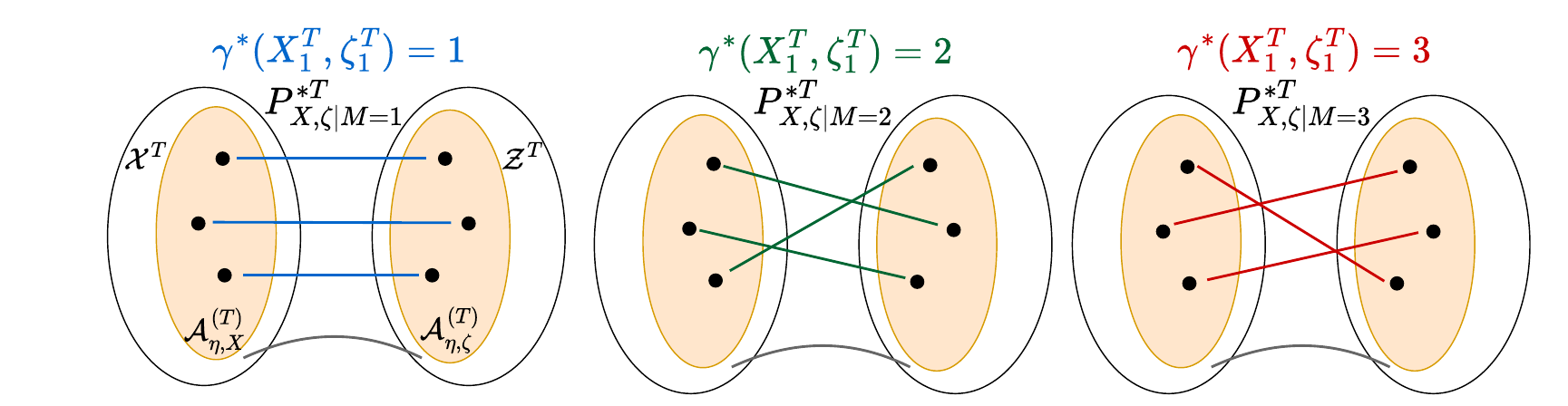}
    \vspace*{-.05in}
    \caption{Illustration of the asymptotically optimal watermarking scheme when $m=3$.}
    \label{fig:asymp opt illustration}
    \vspace*{-0.2in}
\end{figure*}
Under any hypothesis $H_j$ and any $P_{X,\zeta|M=j}$, we define the typical sets of sequences $\{(x^T,\zeta^T)\}$.
\begin{definition}[Typical Sets]
    For arbitrarily small $\eta> 0$, 
    define the  typical sets $\calA_{\eta,X}^{(T)}$ and $\calA_{\eta,\zeta}^{(T)}$ as
    \begin{align}
        \calA_{\eta,X}^{(T)}&\coloneqq \!\bigg\{x^T \!\! \in \calX^T \!\!: \bigg|\frac{1}{T}\frac{1}{\log P_X^T(x^T)}-\sH(P_X) \bigg| \!\! \leq \eta \!\bigg\},\\
        \calA_{\eta,\zeta}^{(T)}&\coloneqq\bigg\{\zeta^T \!\! \in \calZ^T \!\!:  \bigg|\frac{1}{T}\log \frac{1}{P_\zeta^T(\zeta^T)}-\sH(P_\zeta) \!\bigg|\leq \eta\bigg\}.
    \end{align}
\end{definition}

The typical sequences in $\calA_{\eta,X}^{(T)}$ and $\calA_{\eta,\zeta}^{(T)}$ are nearly uniformly distributed and can be mapped with almost deterministic precision. Leveraging the asymptotic equipartition property (AEP), we first present the optimal design when distortion $d=0$ as follows. Here, we use $\doteq$ to denote equality to the first order in the exponent.

\begin{theorem}[Asymptotically Optimal Distortion-Free Watermarking Scheme] \label{Thm:asymp opt}

Let $P_X^*=Q_X$, $\calZ\subset \bbZ$ and design $P_\zeta^*\in\calP(\calZ)$ such that $\sH(P_\zeta^*)=\sH(P_X^*)$. If the number of message bits satisfies $\frac{1}{T}(\log m-\log \alpha)\leq \sH(P_X^*)$, the asymptotically optimal distortion-free watermarking encoders and decoders exist, given as follows.    

Let $\eta=T^{-\frac{1}{4}}$.
The class of optimal decoders is given by
    \setlength{\abovedisplayskip}{5pt}
    \setlength{\belowdisplayskip}{-1pt}
    \setlength{\abovedisplayshortskip}{5pt}
    \setlength{\belowdisplayshortskip}{-1pt}
\begin{align}
    &\Gamma_\eta^*\coloneqq \!\!\Bigg\{ \!\!\gamma \Bigg| 
    \gamma(x_1^T,\zeta_1^T)\!=\!\!\begin{cases}
        g(x_1^T,\zeta_1^T), &\!\!\! \forall x_1^T\in \calA_{\eta,X}^{(T)}, \zeta_1^T \in \calA_{\eta,\zeta}^{(T)}, \\
        &\!\!\! \text{ and } g(x^T,\zeta^T)\leq m;\\
        0, &\text{otherwise},
        \end{cases} \\
    &\begin{array}{l}
    \hspace{-6pt}
         \text{\shortstack{for some function $g:\calA_{\eta,X}^{(T)} \times \calA_{\eta,\zeta}^{(T)}\to \Big[|\calA_{\eta,X}^{(T)}|\Big]$ 
         such that \\ $g(x^T,\cdot)$ and $g(\cdot,\zeta^T)$ are bijective,
         for any fixed $x^T$ and $\zeta^T$.}}
    \end{array}\hspace{-8pt}\Bigg\}
\end{align}
For any $\gamma^*\in\Gamma_\eta^*$, the corresponding asymptotically optimal encoder $f^*$ outputs $P_{X,\zeta|M}^*$ as follows: for any $i\in[m]$, 
\begin{itemize}[leftmargin=*]
    \item for all $x^T\in \calA_{\eta,X}^{(T)}$, 
    $P_{X,\zeta|M}^{*T}(x^T,\zeta^T|i)=\begin{cases}P_X^{*T}(x^T) \doteq e^{-T\sH(P_\zeta^*)}, & \!\!\!\! \text{if $\zeta^T \!\!\in \!\! \calA_{\eta,\zeta}^{(T)}$ and $\gamma^*(x^T,\zeta^T)= i$};\\
    0, & \!\!\!\!\text{otherwise.} 
    \end{cases}$

    \item for all $x^T\notin \calA_{\eta,X}^{(T)}$, let $P_{X,\zeta|X,M}^{*T}(x^T,\zeta^T|i)$  take any non-negative value as long as $\sum_{x^T,\zeta^T}P_{X,\zeta|X,M}^{*T}(x^T,\zeta^T|i)=1$.
\end{itemize}
Thus, for any $\gamma^*\in\Gamma_\eta^*$ and its corresponding $P_{X,\zeta|M}^*$, as $T\to\infty$, we have for all $j\in[m]$,
\begin{align}
    \beta_j(\gamma^*, P_{X,\zeta|M=j}^*)&\leq \exp(-\Omega(T^{\frac{1}{2}}))\to 0, \\
    \text{ and } \sup_{Q_X}\beta_0(\gamma^*, Q_X^T\otimes P_\zeta^{*T})&\leq\alpha+\exp(-\Omega(T^{\frac{1}{2}}))\to \alpha.
\end{align}
\end{theorem}

The proof of Theorem \ref{Thm:asymp opt} is provided in  \cite[Appendix C]{NewMultibitsupp}. The asymptotically optimal decoder deterministically maps a typical sequence $x^T$ to a typical sequence $\zeta^T$ uniquely under different messages $M$. The corresponding optimal joint distribution output by the encoder $f^*$ assigns probability $1$ to such pair of sequences $(x^T,\zeta^T)$, making sure that the detection accuracy is high. Figure \ref{fig:asymp opt illustration} illustrates the design using a toy example when $m=3$. 
\begin{remark}[Existence of $g$ function and implementations]\label{Rmk: existence of g}
    As $|\calA_{\eta,X}^{(T)}|\doteq |\calA_{\eta,\zeta}^{(T)}|$, any Latin square operation \cite{keedwell2015latin} can serve as a valid $g$ function. Below are two examples.
    \begin{itemize}[leftmargin=*]
        \item \underline{Example 1:} Let $n=|\calA_{\eta,X}^{(T)}|$ and index the typical sequences  as $\{(x^T)_i\}_{i=1}^{n}, \{(\zeta^T)_i\}_{i=1}^{n}$. One can define $g((x^T)_i,(\zeta^T)_{(i+M-2) \mod n +1})=M$, for  any $i,M\in [n]$, which takes cyclic permutation of $\calA_{\eta,\zeta}^{(T)}$ as input (as shown in Figure \ref{fig:asymp opt illustration}). 

        \item \underline{Example 2:} Consider the typical sets and the message set as finite fields, $\calA_{\eta,X}^{(T)}=\calA_{\eta,\zeta}^{(T)}=\bbF_q^T$, for some valid $q$. One can define $g(x^T,\zeta^T)=(x^T + \zeta^T) \mod q$.
    \end{itemize}
    
     In general, 
the optimal design can be implemented by lossless coding schemes where the presence of side information $\zeta^T$ ensures that a codeword $X^T$ can be uniquely decoded to one message, e.g., a conditional version of arithmetic coding.
\end{remark}

The information rate of this distortion-free $(m,T)$ watermarking scheme is at most
    \setlength{\abovedisplayskip}{5pt}
    \setlength{\belowdisplayskip}{0pt}
    \setlength{\abovedisplayshortskip}{5pt}
    \setlength{\belowdisplayshortskip}{0pt}
$$R\leq  \sH(Q_X)+\frac{\log\alpha}{T} \xrightarrow{T\to\infty} \sH(Q_X),$$
which achieves the maximum rate in Lemma \ref{Lem: best rate} for  $d=0$.

When we allow some distortion $d>0$ in the watermarking scheme, in Theorem \ref{Thm:asymp opt}, we can change $P_X^*$ to any $P_X$ satisfying $\sD(P_X^T,Q_X^T)\leq d$, and the one that maximizes the information rate is
    \setlength{\abovedisplayskip}{5pt}
    \setlength{\belowdisplayskip}{1pt}
    \setlength{\abovedisplayshortskip}{5pt}
    \setlength{\belowdisplayshortskip}{1pt}
\begin{equation}
    P_X^{*}=\argmax_{P_X:\sD(P_X^T,Q_X^T)\leq d}\sH(P_X).
\end{equation}
When the distortion metric is set as $\sD_\KL$, the optimal  $P_X^{*}$ is the tilting  of $Q_X$ as shown in \cite[Theorem 1]{huang2024odstega}.


Notably, the asymptotic results derived for the i.i.d.~case using classical typical set analysis can be extended to the case where $X^T,\zeta^T$ are stationary ergodic processes. In this generalization, the entropy $H(P_X^*)$ is replaced by the entropy rate of the stationary ergodic process.

\section{Finite-Length Analysis}
Inspired by the asymptotically optimal design, we are now ready to proceed with our analysis in the finite-length setting under the practical non-\iid scenarios. 

We consider the following optimization problem. We aim to minimize the maximum  $j$-th error probability by jointly optimizing the watermarking encoder and decoder, subject to the following constraints: 1) the worst-case false alarm error under control, and 2) the distortion remains  bounded:
    \setlength{\abovedisplayskip}{5pt}
    \setlength{\belowdisplayskip}{0pt}
    \setlength{\abovedisplayshortskip}{5pt}
    \setlength{\belowdisplayshortskip}{0pt}
\begin{align}
    &\min_{\gamma, (\bbP_k)_{k\in[m]}} \max_{j\in[m]}  \; \beta_j(\gamma, \bbP_j)  \tag{P1} \label{Eq: opt-O}
    \\
     &\qquad \text{s.t.}  \; \sup_{Q_{X^T}} \beta_0(\gamma,Q_{X^T}\otimes P_{\zeta^T})\leq \alpha, \quad \sD(P_{X^T},Q_{X^T})\leq d.
\end{align}
Recall that $\bbP_k\coloneqq P_{X^T,\zeta^T|M=k}$. Assumption \ref{Ass: independence} implicitly imposes the constraint that all $(\bbP_k)_{k\in[m]}$ share the same marginals projected on $\calX^T$ and  $\calZ^T$. 

The following theorem characterizes the min-max $j$-th error probability of this optimization problem---the universal minimax that holds for all watermarking schemes within our framework. The proof is provided in  \cite[Appendix D]{NewMultibitsupp}.
\vspace{-0.5em}
\begin{theorem}[Universal Min-Max $j$-th Error] \label{Thm: univ mim jth error}
The universal min-max $j$-th error, denoted  by $\beta^*(m,T,\alpha,d)$, from \eqref{Eq: opt-O} is 
    \setlength{\abovedisplayskip}{5pt}
    \setlength{\belowdisplayskip}{5pt}
    \setlength{\abovedisplayshortskip}{5pt}
    \setlength{\belowdisplayshortskip}{5pt}
\begin{align}
   \hspace{-1em}\beta^*(m,T,\alpha,d)=\hspace{-1em}\min_{P_{X^T}: \sD(P_{X^T},Q_{X^T})\leq d} \sum_{x^T} \!\bigg(\! P_{X^T}(x^T)-\frac{\alpha}{m} \bigg)_+ \!\!\!. \label{Eq:univ mim-max jth error}
\end{align}
\end{theorem}
\vspace{-.5em}
In the converse proof, we first fix any decoder $\gamma$ and show a lower bound for the min-max $j$-th error probability, i.e., \eqref{Eq:univ mim-max jth error}. The lower bound is independent of  $\gamma$ and thus holds for the optimal value of \eqref{Eq: opt-O}. In the achievability proof, we present a watermarking scheme that achieves this lower bound, as stated in the next theorem. 
We observe that $\beta^*(m,T,\alpha,d)$ resembles the universally minimum Type-II error for zero-bit watermarking schemes \cite{he2024universally}, with $\alpha$ replaced by $\frac{\alpha}{m}$. As the message set size $m$ increases (with $\alpha$ and $T$ fixed), $\beta^*(m,T,\alpha,d)$ increases to $1$. Conversely, increasing the false alarm threshold $\alpha$ or the token length $T$ allows for a larger message set $m$ while keeping $\beta^*(m, T, \alpha,d)$ under control. If $m\leq \alpha|\calX|^T$, there exist cases where $\beta^*(m, T, \alpha,d)=0$. Hence, Theorem \ref{Thm: univ mim jth error} reveals a fundamental trade-off between detectability, token length and  message size. Moreover, as the distortion $d$ increases, the lower bound for $\beta_j^*$ decreases. This means that a watermarking scheme can trade off text quality for lower detection errors. 



The following theorem presents a class of $d$-distorted $(m,T)$ watermarking schemes for $m\leq |\calX|^T$ that satisfies Assumption \ref{Ass: independence} and achieves Theorem \ref{Thm: univ mim jth error}.
\vspace{-0.5em}
\begin{theorem}[Optimal $d$-Distorted Watermarking Scheme]\label{Thm: finite opt}
    Choose $\calZ^T\subset \bbZ^T$ such that $|\calZ|^T=|\calX|^T+1$. Randomly pick a redundant sequence  $\tilde{\zeta^T}\in\calZ^T$.
If the message set size $m\leq |\calX|^T$, the optimal watermarking encoders and decoders that achieve Theorem \ref{Thm: univ mim jth error} exist, given as follows.

The class of optimal decoders is given by
\begin{small}
    \begin{align}
    &\Gamma_{\tilde{\zeta^T}}^*\hspace{-3pt}\coloneqq \hspace{-3pt} \Bigg\{\gamma \Bigg| 
    \gamma(x^T,\zeta^T)=\begin{cases}
        h(x^T,\zeta^T), &\text{if }  \zeta^T\ne \tilde{\zeta^T} \text{ and }  \\
         &   h(x^T,\zeta^T)\leq m,\\
        0, &\text{otherwise},
    \end{cases}\\
    &\hspace{-8pt}\begin{array}{l}\text{\shortstack[l]{ for some function $h:\calX^T \times \calZ^T\backslash{\{\tilde{\zeta^T}\}} \to [|\calX^T|]$  such that \\ $h(x^T,\cdot)$ and $h(\cdot, \zeta_1^T)$  are bijective, given any fixed $x^T$ \!\!\! and $\zeta_1^T$.}} \end{array}\hspace{-10pt} \Bigg\}.
\end{align}
\end{small}
For any $\gamma^*\in \Gamma_{\tilde{\zeta^T}}^*$, the corresponding optimal encoder $f^*$ outputs $\bbP_j^*$ as follows: for any $j\in[m]$, $\bbP_j^*(x^T,\zeta^T)=$
\begin{small}
   \begin{equation}
   \!\!\! 
    \begin{aligned}
        \begin{cases}
        P^*_{X^T}(x^T) \wedge P_{\zeta^T}^*(\zeta^T), \qquad \text{if }\; \gamma^*(x^T,\zeta^T)=j;\\
        \!\! \frac{\big(P^*_{X^T}(x^T)-P_{\zeta^T}^*(\gamma_j^{*-1}(x^T))\big)_+ \cdot \big(P_{\zeta^T}^*(\zeta^T)-P^*_{X^T}(\gamma_j^{*-1}(\zeta^T))\big)_+}{\beta^*(m,T,\alpha,d)}, \\
        \hspace{6cm} \text{otherwise} ,
    \end{cases}
    \end{aligned} 
    \label{Eq: minmax multibit scheme}
    \end{equation}
\end{small}
    where $\gamma_j^{*-1}$ represents the inverse of $\gamma^*$ for a fixed $j\in[m]$ ($\gamma_j^{*-1}(\tilde{\zeta^T})=\emptyset$ in particular), 
    $$
    P_{X^T}^*=\argmin_{\substack{P_{X^T}:\sD(P_{X^T} , Q_{X^T} )\leq d
    }} \sum_{x^T}\bigg(P_{X^T}(x^T)- \frac{\alpha}{m} \bigg)_+, \text{ and}$$
    \begin{small}
    $$
    P_{\zeta^T}^*=\bigg(\!\!\underbrace{\bigg(P_{X^T}^*(x^T)\wedge \frac{\alpha}{m}\bigg)_{x^T\in\calX^T}}_{P_{\zeta^T}^*(\zeta^T), \; \forall \zeta^T\in \calZ^T\backslash{\{\tilde{\zeta^T}\}} }, \underbrace{\sum_{x^T\in\calX^T}\!\!\!\bigg(P_{X^T}^*(x^T)- \frac{\alpha}{m} \bigg)_{\!\!+}}_{P_{\zeta^T}^*(\tilde{\zeta^T})} \!\bigg).$$
    \end{small}
\end{theorem}

\vspace{-6pt}
The proof of Theorem \ref{Thm: finite opt} is provided in the achievability part in  \cite[Appendix D]{NewMultibitsupp}. This optimal watermarking scheme ensures the secrecy of embedded message, i.e., $P^*_{X^T|M}=P^*_{X^T}$ and $P^*_{\zeta^T|M}=P^*_{\zeta^T}$. The construction of the encoder's output distributions $(\bbP_j^*)_{j\in[m]}$ are adaptive to the original LLM output $Q_{X^T}$, which can regarded as an extension of \cite[Theorem 2]{he2024universally}. It is equivalent to transporting the probability mass from $\calV^T$ to $\calZ^T$, maximizing $\bbP_M^*(x^T,\zeta^T)$ for $\gamma(x^T,\zeta^T)=M$, while keeping the worst-case false alarm error below $\alpha$. 
Moreover, the introduction of $\tilde{\zeta^T}$ helps to control the worst-case false alarm. If $P_{X^T}^*(x^T)>\frac{\alpha}{m}$ (i.e., low-entropy text), $x^T$ may be mapped to $\tilde{\zeta^T}$ during watermarking, which makes it harder to detect as watermarked. In conclusion, the proposed scheme provides a structured approach to improving the implementation of multi-bit LLM watermarking.



\section{Discussion and Future Works}

While our theoretical analysis of the distributional information embedding problem does not fully account for all aspects of LLMs (e.g., auto-regressive nature), we believe it provides valuable insights for designing multi-bit watermarking schemes. We rigorously demonstrate that the best achievable rate in the asymptotic regime is determined by the entropy of the text distribution  $\sH(P_X)$, establishing a fundamental limit that serves as a benchmark for evaluating existing multi-bit watermarking schemes.

Moreover, this result implies an inherent connection between the problem of distributional information embedding and lossless compression, where the fundamental limit is also the entropy of the source distribution. Interestingly, \cite{huang2024odstega} proposes a steganography algorithm that exploits this connection by using the decoder of an arithmetic coding scheme\footnote{The decoder of source coding maps message bit to symbols (tokens), and the encoder maps tokens to message.} 
as the LLM watermarking encoder to sample from tokens while employing the arithmetic coding encoder as the decoder in our context. This duality between the two problems suggests that new watermarking schemes could be inspired by existing source coding techniques, presenting an intriguing direction for future exploration.

\bibliographystyle{IEEEtran}
\bibliography{sample, IT_references}

\onecolumn
\appendix

\subsection{Proof of Lemma \ref{Lem: best rate}}\label{App: pf of best rate}
\begin{proof}
Let $P_e=\Pr(\hatM\ne M)$. 
From the Fano's inequality, we have
\begin{equation}
    \sH(M|\hatM,\zeta^T)\leq \sH(M|\hatM)\leq 1 +P_e\log m.
\end{equation}
The entropy of $M$ is upper bounded by
\begin{align}
    \log m = \sH(M)=\sH(M|\zeta^T)&=\sI(M;\hatM|\zeta^T)+\sH(M|\hatM,\zeta^T)\\
    &\leq \sI(M;X^T|\zeta^T)+1 +P_e\log m\\
    &\leq H(X^T|\zeta^T)+1 +P_e\log m,
\end{align}
which leads to
\begin{align}
    \frac{\log m}{T}\leq \frac{H(X^T|\zeta^T)}{T}+\frac{1}{T}+P_e\frac{\log m}{T}.
\end{align}
If $P_e\to 0$ as $T\to\infty$, we have
\begin{align}
    \frac{\log m}{T} \leq \frac{H(X^T|\zeta^T)}{T}\leq \sH(P_X)\leq \sup_{P_X:\sD(P_X^T,Q_X^T)\leq d}\sH(P_X).
\end{align}

\end{proof}

\subsection{Proof of Lemma \ref{Lem:LB for j-th error}}\label{App: pf of Lem:LB for j-th error}

\begin{proof}
    For any $i\ne j$, define the relative entropy typical set
    \begin{align}
        \calA_{\epsilon,i,j}^{(T)}(\bbP_i\|\bbP_j)\coloneqq \bigg\{ (x^T,\zeta^T):\bigg|\frac{1}{T}\log \frac{\bbP_i(x^T,\zeta^T)}{\bbP_j(x^T,\zeta^T)}-\sD_\KL(P_{X,\zeta|M=i}\|P_{X,\zeta|M=j}) \bigg|\leq \epsilon \bigg\}.
    \end{align}
    We have $\bbP_j(\calB_{T,j}^\rmc)=1-\bbP_j(\calB_{T,j})$ and
    \begin{align}
        \bbP_j(\calB_{T,j})&=1-\sum_{i:i\ne j}\bbP_j(\calB_{T,i}) \leq 1-\sum_{i:i\ne j}\bbP_j(\calB_{T,i}\cap \calA_{\epsilon,i,j}^{(T)} )\\
        &\leq 1-\sum_{i:i\ne j}\sum_{(x^T,\zeta^T)\in \calB_{T,i}\cap \calA_{\epsilon,i,j}^{(T)} } \bbP_i(x^T,\zeta^T)\exp(-T(\sD_\KL(P_{X,\zeta|M=i}\|P_{X,\zeta|M=j})+\epsilon))\\
        &=1-\sum_{i:i\ne j}\exp(-T(\sD_\KL(P_{X,\zeta|M=i}\|P_{X,\zeta|M=j})+\epsilon))\bbP_i(\calB_{T,i}\cap \calA_{\epsilon,i,j}^{(T)} )\\
        &\overset{\text{(a)}}{\leq} 1-\sum_{i:i\ne j}\exp(-T(\sD_\KL(P_{X,\zeta|M=i}\|P_{X,\zeta|M=j})+\epsilon))(1-2\epsilon)\\
        &\leq 1-m(1-2\epsilon)\exp(-T(\min_{i:i\ne j}\sD_\KL(P_{X,\zeta|M=i}\|P_{X,\zeta|M=j})+\epsilon))\\
        &\leq 1-m(1-2\epsilon)\exp(-T(\max_{P_X:\sD(P_X^T,Q_X^T)\leq d}\min_{i:i\ne j}\sD_\KL(P_{X,\zeta|M=i}\|P_{X,\zeta|M=j})+\epsilon))
    \end{align}
    where (a) follows since $\bbP_i(\calB_{T,i}\cap \calA_{\epsilon,i,j}^{(T)} )=1-\bbP_i(\calB_{T,i}^\rmc\cup (\calA_{\epsilon,i,j}^{(T)})^\rmc )\geq 1-\bbP_i(\calB_{T,i}^\rmc)-\bbP_i((\calA_{\epsilon,i,j}^{(T)})^\rmc ) \geq 1-2\epsilon$ for sufficiently large $T$. The proof is thus complete.
\end{proof}

\subsection{Proof of Theorem \ref{Thm:asymp opt}}\label{App: pf of Thm:asymp opt}
\paragraph{Existence of asymptotically optimal decoders}
First, the function $g$ proposed in Theorem \ref{Thm:asymp opt} always exists, as discussed in Remark \ref{Rmk: existence of g}.
If the number of message bits satisfies $\frac{1}{T}(\log m-\log \alpha)\leq \sH(P_X^*)$, then we have
\begin{equation}
    m \dotleq e^{T \sH(P_X^*)} \doteq \calA_{\eta,X}^{(T)},
\end{equation}
and the output space of $g$ contains $[m]$. Thus, any decoder in the  class of asymptotically optimal decoders $\Gamma_\eta^*$ can decode messages drawn from $[m]$.

\paragraph{Asymptotic optimality} 
For any $\gamma\in \Gamma_\eta^*$, one can always construct the corresponding encoder outputs $P_{X,\zeta|M}^*$ in Theorem \ref{Thm:asymp opt}. In the following, we first show that the probability of the atypical set decays exponentially with $T$. We then prove that the $j$-th error probability vanishes to $0$ while the worst-case false alarm error is upper bounded by $\alpha$ as $T\to\infty$.

Let $\eta=T^{-\frac{1}{4}}$ and define the set $\calA_{\eta,j}^{(T)}$ of jointly typical sequences $\{(x^T,\zeta^T)\}$ w.r.t.\ the distribution  $P_{X,\zeta|M=j}$ as
    \begin{align}
        \calA_{\eta,j}^{(T)}\coloneqq\bigg\{(x^T,\zeta^T)\in \calX^T\times \calZ^T: &\bigg|-\frac{1}{T}\log P_X^T(x^T)- \sH(P_X)  \bigg| \leq \eta, \bigg| -\frac{1}{T}\log P_\zeta^T(\zeta^T)- \sH(P_\zeta) \bigg| \leq \eta, \nn\\
        & \bigg|-\frac{1}{T}\log P_{X,\zeta|M=j}^T(x^T,\zeta^T)-\sH(P_{X,\zeta|M=j}) \bigg|\leq \eta \bigg\}.
    \end{align}

First, we bound the probability of the atypical sets $(\calA_{\eta,X}^{(T)})^\rmc,(\calA_{\eta,\zeta}^{(T)})^\rmc,(\calA_{\eta,j}^{(T)})^\rmc$.
From the union bound, we have
\begin{align}
    \bbP_j((X^T,\zeta^T)\notin \calA_{\eta,j}^{(T)})&\leq \bbP_j\bigg(\bigg|-\frac{1}{T}\log P_X^T(x^T)- \sH(P_X)  \bigg| \geq  \eta \bigg)+\bbP_j\bigg(\bigg|-\frac{1}{T}\log P_\zeta^T(\zeta^T)- \sH(P_\zeta)  \bigg| \geq  \eta \bigg)\\
    &\quad +\bbP_j\bigg(\bigg|-\frac{1}{T}\log P_{X,\zeta|M=j}^T(x^T,\zeta^T)-\sH(P_{X,\zeta|M=j}) \bigg| \geq  \eta \bigg). \label{Eq: atypical union bd}
\end{align}
Then, by the Chernoff bound, we have
\begin{align}
    \bbP_j\bigg(\bigg|-\frac{1}{T}\log P_X^T(x^T)- \sH(P_X)  \bigg| \geq  \eta \bigg)&\leq 2\bbP_j\bigg(-\frac{1}{T}\log P_X^T(x^T)- \sH(P_X)  \geq  \eta \bigg)\\
    &\leq 2\exp\bigg(-T\sup_{s\geq 0}(s\eta-\log \bbE[\exp(-s\log P_{X^T}(X^T))]) \bigg)\\
    &\overset{\text{(a)}}{\approx} 2\exp\bigg(-T\sup_{s\geq 0}(s\eta-\big(-s\bbE[\log P_{X^T}(X^T)]+s^2\bbE[(\log P_{X^T}(X^T))^2] \big) \bigg)\\
    &\eqb 2\exp(-\Omega(T\eta^2))=\exp(-\Omega(T^{\frac{1}{2}})),
\end{align}
where (a) follows from the Taylor expansion of $\exp(\cdot)$ and $\log(\cdot)$ and (b) follows since the maximum is achieved by $s=O(\eta)$.
The rest of the terms in the union bound \eqref{Eq: atypical union bd} can be similarly proved.

Thus, the probability of the jointly atypical set is upper bounded by
\begin{align}
    \bbP_j((X^T,\zeta^T)\notin \calA_{\eta,j}^{(T)})\leq 3\exp(-\Omega(T^{\frac{1}{2}}))=\exp(-\Omega(T^{\frac{1}{2}})).
\end{align}
    

Next, we prove that the proposed watermarking scheme in Theorem \ref{Thm:asymp opt} achieves the asymptotic optimality.
Let $P_X^*=Q_X$, $\calZ\subset \bbZ$ and design $P^*_\zeta\in\calP(\calZ)$ such that $\sH(P_\zeta^*)=\sH(P_X^*)$.


For any $\gamma^*\in\Gamma^*$, under the watermarking scheme given in Theorem \ref{Thm:asymp opt}, for any $j\in[m]$, the $j$-th error probability is given by
\begin{align}
    \beta_j(\gamma^*,P^*_{X^T,\zeta^T|M=j})&=\sum_{x^T,\zeta^T} P^*_{X^T,\zeta^T|M}(x^T,\zeta^T|j)\mathbbm{1}\{\gamma^*(x^T,\zeta^T)\ne j\}\\
    &\leq \sum_{(x^T,\zeta^T)\in \calA_{\eta,j}^{(T)}} P^*_{X^T,\zeta^T|M}(x^T,\zeta^T|j)\mathbbm{1}\{\gamma^*(x^T,\zeta^T)\ne j\} + \exp(-\Omega(T^{\frac{1}{2}}))\\
    &=\exp(-\Omega(T^{\frac{1}{2}})) \to 0 \text{ as } T\to \infty.
\end{align}
For $j=0$, the worst-case false alarm error probability is upper bounded as follows. For any $x^T\in \calA_{\eta,X}^{(T)}$,
\begin{align}
    \sum_{\zeta^T}P^*_\zeta(\zeta^T)\mathbbm{1}\{\gamma^*(x^T,\zeta^T)\ne 0\}&\leq \sum_{\zeta^T\in\calA_{n,\zeta}^{(T)}}P^*_\zeta(\zeta^T)\mathbbm{1}\{\gamma^*(x^T,\zeta^T)\ne 0\} +\exp(-\Omega(T^{\frac{1}{2}}))\\
    &=\sum_{i\in[m]}\sum_{\zeta^T\in\calA_{n,\zeta}^{(T)}}P^*_\zeta(\zeta^T)\mathbbm{1}\{\gamma^*(x^T,\zeta^T)=i\} +\exp(-\Omega(T^{\frac{1}{2}}))\\
    &= \sum_{i\in [m]}\sum_{\zeta^T\in\calA_{n,\zeta}^{(T)}} \bigg(\frac{1}{m}\sum_{j\in[m]} \sum_{x^T}P^*_{X^T,\zeta^T|M}(x^T,\zeta^T|j) \bigg)\mathbbm{1}\{\gamma^*(x^T,\zeta^T)=i\} +\exp(-\Omega(T^{\frac{1}{2}}))\\
    &\doteq \sum_{i\in [m]}\sum_{\zeta^T\in\calA_{n,\zeta}^{(T)}}e^{-T\sH(\zeta)}\mathbbm{1}\{\gamma^*(x^T,\zeta^T)=i\} +\exp(-\Omega(T^{\frac{1}{2}}))\\
    &= m e^{-T\sH(\zeta)}+\exp(-\Omega(T^{\frac{1}{2}})) \\
    &\overset{\text{(a)}}{\leq} \alpha+\exp(-\Omega(T^{\frac{1}{2}}))\\
    &\xrightarrow{T\to\infty}\alpha,
\end{align}
where (a) follows from the condition $\log m \leq \log \alpha+T\rmH(P_\zeta^*)$ in Theorem \ref{Thm:asymp opt}.

For any $x^T\in (\calA_{\eta,X}^{(T)})^\rmc$,
\begin{align}
    \sum_{\zeta^T}P^*_\zeta(\zeta^T)\mathbbm{1}\{\gamma^*(x^T,\zeta^T)\ne 0\}=0.
\end{align}
Since any distribution $Q_{X}^T$ can be written as a linear combinations of $\{\delta_{x^T}\}_{x^T\in\calX^T}$, we have
\begin{align}
    \sup_{Q_X}\beta_0(\gamma^*, Q_X \otimes P_\zeta^* )=\sup_{Q_X}\sum_{x^T,\zeta^T} Q_X^T(x^T)P^*_\zeta(\zeta^T)\mathbbm{1}\{\gamma^*(x^T,\zeta^T)\ne 0\}\to \alpha, \text{ as } T\to\infty.
\end{align}

\subsection{Proof of Theorem \ref{Thm: univ mim jth error} and Theorem \ref{Thm: finite opt}}\label{App: pf of min jth error}

We restate the optimization problem \eqref{Eq: opt-O} as follows:
\begin{align}
    &\min_{\gamma,\bbP_1,\ldots, \bbP_m} \max_{j\in[m]}\;\beta_j(\gamma, P_{X^T,\zeta^T|M=j})
    \\
    &\qquad \text{s.t.} \quad \sup_{Q_{X^T}} \beta_0(\gamma,Q_{X^T}\otimes P_{\zeta^T})\leq \alpha, \\
    &\qquad \qquad \; \sD(P_{X^T},Q_{X^T})\leq d.
\end{align}
Assumption \ref{Ass: independence} implicitly imposes the constraint that all $\bbP_1,\ldots, \bbP_m$ should have the same marginal distributions projected on $\calX^T$ and on $\calZ^T$, i.e., $P_{X^T}$  and $P_{\zeta^T}$.

\paragraph{Converse}
\begin{proof}[Proof of lower bound] First, let us fix a decoder $\gamma$.
    From the worst-case false alarm constraint, we have
    \begin{align}
        \alpha\geq \sup_{Q_{X^T}} \beta_0(\gamma,Q_{X^T}\otimes P_{\zeta^T})&\geq \sum_{\zeta^T}P_{\zeta^T}(\zeta^T)\mathbbm{1}\{\gamma(x^T,\zeta^T)\ne 0\}\\
        &=\sum_{i\in[m]}\sum_{\zeta^T}P_{\zeta^T}(\zeta^T)\mathbbm{1}\{\gamma(x^T,\zeta^T)=i\}, \quad \forall x^T.
    \end{align}
    Therefore, we have
    \begin{align}
        \sum_{\zeta^T}P_{\zeta^T}(\zeta^T)\mathbbm{1}\{\gamma(x^T,\zeta^T)=i\}\leq\alpha_i, \quad \forall i, x^T, \quad \sum_{i\in[m]}\alpha_i=\alpha. \label{Eq: ith FA constraint}
    \end{align}

    The $j$-th error probability is lower bounded by
    \begin{align}
        \beta_j(\gamma,P_{X^T,\zeta^T|M=j})&=1-\bbP_j(\gamma(X^T,\zeta^T)=j)\\
        &=1-\sum_{x^T,\zeta^T}P_{\zeta^T}(\zeta^T)P_{X^T|\zeta^T,M}(x^T|\zeta^T,j)\mathbbm{1}\{\gamma(x^T,\zeta^T)=j\}.
    \end{align}

From \eqref{Eq: ith FA constraint}, we have
\begin{align}
    \sum_{\zeta^T}P_{\zeta^T}(\zeta^T)P_{X^T|\zeta^T,M}(x^T|\zeta^T,j)\mathbbm{1}\{\gamma(x^T,\zeta^T)=j\}\leq \sum_{\zeta^T}P_{\zeta^T}(\zeta^T)\mathbbm{1}\{\gamma(x^T,\zeta^T)=j\}\leq \alpha_j,
\end{align}
and since $\mathbbm{1}\{\gamma(x^T,\zeta^T)=j\}\leq 1$,
\begin{align}
    \sum_{\zeta^T}P_{\zeta^T}(\zeta^T)P_{X^T|\zeta^T,M}(x^T|\zeta^T,j)\mathbbm{1}\{\gamma(x^T,\zeta^T)=j\}\leq \sum_{\zeta^T}P_{\zeta^T}(\zeta^T)P_{X^T|\zeta^T,M}(x^T|\zeta^T,j)=P_{X^T}(x^T).
\end{align}
Therefore, we lower bound $\beta_j$ as follows
\begin{align}
    \beta_j(\gamma,P_{X^T,\zeta^T|M=j})&\geq  1- \sum_{x^T} (P_{X^T}(x^T)\wedge \alpha_j)=\sum_{x^T} (P_{X^T}(x^T)-\alpha_j)_+\\
    &\geq \min_{P_{X^T}: \sD(P_{X^T},Q_{X^T})\leq d}\sum_{x^T} (P_{X^T}(x^T)-\alpha_j)_+ \quad \forall j\in[m].
\end{align}
Among all possible $(\alpha_1,\ldots, \alpha_m)$ that sum up to $\alpha$, the vector that minimizes the lower bound for $\max_{j\in[m]}\beta_j(\gamma,P_{X^T,\zeta^T|M=j})$ is $(\frac{\alpha}{m},\ldots, \frac{\alpha}{m})$. The proof is as follows:
\begin{align}
    \max_{j\in[m]}\beta_j(\gamma,P_{X^T,\zeta^T|M=j})&\geq \max_{j\in[m]}\min_{P_{X^T}: \sD(P_{X^T},Q_{X^T})}\sum_{x^T} (P_{X^T}(x^T)-\alpha_j)_+ \\
    &\overset{(a)}{\geq} \min_{P_{X^T}: \sD(P_{X^T},Q_{X^T})\leq d} \sum_{x^T} \bigg(P_{X^T}(x^T)-\frac{\alpha}{m} \bigg)_+, \label{Eq: lower bd for max beta_j}
\end{align}
where (a) holds with equality when $\alpha_j=\frac{\alpha}{m}$ for all $j\in[m]$. 

We observe that the lower bound \eqref{Eq: lower bd for max beta_j} is independent of $\gamma$. Thus, the lower bound also holds for the optimal value of the optimization problem \eqref{Eq: opt-O}.

\end{proof}

\paragraph{Achievability}
Choose $\calZ^T\subset \bbZ^T$ such that $|\calZ|^T=|\calX|^T+1$. Randomly pick a redundant sequence  $\tilde{\zeta^T}\in\calZ^T$.
For any $m\leq |\calX|^T$, define a set of decoders as
    \begin{align}
    &\Gamma_{\tilde{\zeta^T}}^*\hspace{-3pt}\coloneqq \hspace{-3pt} \left\{\gamma \left| \begin{array}{c}
    \gamma(x^T,\zeta^T)=\begin{cases}
        j, &\text{if }  \zeta^T\ne \tilde{\zeta^T}  \\
         &\hspace{-5pt} \text{and }  h(x^T,\zeta^T)=j\leq m,\\
        0, &\text{otherwise},
    \end{cases},\\
    \text{\shortstack{ for some function $h:\calX^T \times \calZ^T\backslash{\{\tilde{\zeta^T}\}} \to [|\calX^T|]$ satisfying that \\ $h(x^T,\cdot)$ and $h(\cdot, \zeta_1^T)$ are both bijective, given any fixed $x^T$ and fixed $\zeta_1^T$.}} \end{array}\right. \hspace{-10pt}\right\}.
\end{align}

Construct $P_{\zeta^T|M}^*=P_{\zeta^T}^*$ as follows
\begin{align}
    P_{\zeta^T}^*=\left(\underbrace{\bigg(P_{X^T}^*(x^T)\wedge \frac{\alpha}{m}\bigg)_{x^T\in\calX^T}}_{P_{\zeta^T}^*(\zeta^T), \; \forall \zeta^T\in \calZ^T\backslash{\{\tilde{\zeta^T}\}} }, \underbrace{\sum_{x^T\in\calX^T}\bigg(P_{X^T}^*(x^T)- \frac{\alpha}{m} \bigg)_+}_{P_{\zeta^T}^*(\tilde{\zeta^T})} \right) \in \calP(\calZ^T),
\end{align}
where $P_{\zeta^T}^*(\tilde{\zeta^T})=\sum_{x^T\in\calX^T}\bigg(P_{X^T}^*(x^T)- \frac{\alpha}{m} \bigg)_+$. 

In particular, if we choose the support as  $\calZ^T=\calX^T\cup \{\tilde{\zeta^T}\}$, the total variation distance between any $P_{X^T}$ and $P_{\zeta^T}^*$ is 
\begin{align}
    \sD_\TV(P_{X^T},P_{\zeta^T}^*)=\sum_{x^T\in\calX^T}\bigg(P_{X^T}(x^T)- \frac{\alpha}{m} \bigg)_+. \label{Eq: TV between X and zeta}
\end{align}
In the following, with no risk of confusion, we will refer to $\sD_\TV(P_{X^T},P_{\zeta^T}^*)$ as the quantity defined in \eqref{Eq: TV between X and zeta}, even if a different support $\calZ^T$ is chosen.

Construct $\bbP_j^*=P_{X^T,\zeta^T|M=j}^*$ as follows,
    \begin{equation}
    \begin{aligned}
        P^*_{X^T,\zeta^T|M=j}(x^T,\zeta^T)=\begin{cases}
        P^*_{X^T}(x^T) \wedge P_{\zeta^T}^*(\zeta^T), &\text{if}\; \gamma^*(x^T,\zeta^T)=j;\\
        \frac{\big(P^*_{X^T}(x^T)-P_{\zeta^T}^*(\gamma_j^{*-1}(x^T))\big)_+ \cdot \big(P_{\zeta^T}^*(\zeta^T)-P^*_{X^T}(\gamma_j^{*-1}(\zeta^T))\big)_+}{\sD_\TV(P^*_{X^T},P_{\zeta^T}^*)}, &\text{otherwise},
    \end{cases}
    \end{aligned} 
    \label{Eq: minmax multibit scheme}
    \end{equation}
    where $\gamma_j^{*-1}$ represents the inverse of $\gamma^*$ for a fixed $j\in[m]$ ($\gamma_j^{*-1}(\tilde{\zeta^T})=\emptyset$ in particular) and
    $$
    P_{X^T}^*
    = \argmin_{\substack{P_{X^T}:\sD(P_{X^T}, Q_{X^T} )\leq d
    }} \sD_\TV(P_{X^T},P_{\zeta^T}^*)=\argmin_{\substack{P_{X^T}:\sD(P_{X^T} , Q_{X^T} )\leq d
    }} \sum_{x^T\in\calX^T}\bigg(P_{X^T}(x^T)- \frac{\alpha}{m} \bigg)_+.$$
    
This conditional joint distribution $P^*_{X^T,\zeta^T|M=j}$ with fixed marginals minimizes the $j$-th error probability $\bbP_j(\gamma^*(X^T,\zeta^T)\ne j)$, as shown below: 
\begin{align}
    \bbP_j^*(\gamma^*(X^T,\zeta^T)\ne j)&=1-\sum_{x^T,\zeta^T: \gamma^*(x^T,\zeta^T)=j}(P^*_{X^T}(x^T) \wedge P_{\zeta^T}^*(\zeta^T))\\
    &=1-\sum_{x^T,\zeta^T: \gamma^*(x^T,\zeta^T)=j}\bigg(P^*_{X^T}(x^T) \wedge \frac{\alpha}{m} \bigg)\\
    &=\sum_{x^T\in\calX^T}\bigg(P_{X^T}^*(x^T)- \frac{\alpha}{m} \bigg)_+,
\end{align}
and ensures that 
\begin{align}
    \sum_{\zeta^T}P_\zeta^*(\zeta^T)\mathbbm{1}\{\gamma^*(x^T,\zeta^T)\ne 0\}=\sum_{i\in[m]}P_\zeta^*(\zeta^T)\mathbbm{1}\{\gamma^*(x^T,\zeta^T)=i\}\leq m\cdot \frac{\alpha}{m}=\alpha, \quad \forall x^T.
\end{align}

Therefore, the scheme proposed in \eqref{Eq: minmax multibit scheme} achieves the min-max $j$-th error probability in Theorem \ref{Thm: univ mim jth error}.


\end{document}